\documentclass{article}
\usepackage{amssymb}

\usepackage{amsmath}

\input{tcilatex}

\begin{document}

\title{Quasistationary quaternionic Hamiltonians and complex stochastic maps}
\author{G. Scolarici\thanks{%
e-mail: scolarici@le.infn.it} and L. Solombrino\thanks{%
e-mail: solombrino@le.infn.it} \\
Dipartimento di Fisica dell'Universit\`{a} del Salento \\
and INFN, Sezione di Lecce, I-73100 Lecce, Italy}
\maketitle

\begin{abstract}
We show that the complex projections of time-dependent $\eta $%
-quasianti-Hermitian quaternionic Hamiltonian dynamics are complex
stochastic dynamics in the space of complex quasi-Hermitian density matrices
if and only if a quasistationarity condition is fulfilled, i. e., if and
only if $\eta $ is an Hermitian positive time-independent complex operator.
An example is also discussed.
\end{abstract}

\section{\protect\bigskip Introduction}

Studies on non-Hermitian $\mathcal{PT}$-symmetric (or, better,
pseudo-Hermitian) Hamiltonians \cite{proc} proven that it is possible to
formulate a consistent quantum theory based on such non-Hermitian
Hamiltonians \cite{mosta} at least whenever diagonalizable time-independent
Hamiltonians having a real spectrum are taken into account. It was further
shown that if the above hypotheses hold, complex quasi-Hermitian systems can
be described as open systems, and a master equation was derived \cite%
{bologna}, proving that the evolution of such systems obeys a one-parameter
semigroup law.

Moreover, the theory of open quantum systems can be obtained, in many
relevant physical situations, as the complex projection of quaternionic
closed quantum systems \cite{compent}, \cite{Asor}, \cite{gallipoli}, \cite%
{ASS}. In particular, it was shown that the complex projection of $\eta $%
-quasianti-Hermitian quaternionic time-independent Hamiltonian dynamics are
ruled by one-parameter semigroups of maps in the space of complex
quasi-Hermitian density matrices if and only if $\eta $ is an Hermitian
positive \textit{complex} operator \cite{scola}, \cite{gius}.

In this paper, we will go more inside to this subject, by considering
time-dependent Hamiltonians. Such a problem was recently investigated \cite%
{timdep} in the complex case, and a necessary and sufficient condition was
derived for the unitarity of time evolution. We intend here to exploit an
analogous method in the quaternionic case, on one hand; on the other hand,
we will show that complex stochastic maps (in the sense by Sudarshan \textit{%
et al.} \cite{s}) can be obtained by complex projection of time-dependent $%
\eta $-quasistationary quaternionic Hamiltonians.

This paper is organized as follows. In Sec. 2 we firstly recall some
previous results on quaternionic pseudo-Hermitian density matrices and next
we derive a necessary and sufficient condition for the $\eta $-unitarity of
the time evolution associated with a $\eta $-pseudoanti-Hermitian
quaternionic Hamiltonian. In Sec. 3 we restrict ourselves to consider
positive operators $\eta $ only, and we define $\eta $-quasistationary
quaternionic Hamiltonians. The dynamics ruled by such Hamiltonians are then
investigated and explicitly written down, together with their complex
projection which constitue stochastic maps in the space of the $\eta $%
-quasi-Hermitian complex density matrices. Such results are illustrated by
an example in Sec. 4. Finally, in Sec. 5, we show how one can obtain the
full class of $\eta $-quasistationary quaternionic dynamics, and construct
the stochastic map associated with a generic time-dependent anti-Hermitian
quaternionic operator.

\section{Pseudoanti-Hermitian quaternionic Hamiltonian dynamics}

In this section, we will introduce the notion of quaternionic
pseudo-Hermitian density matrix and a corresponding Liouville-von Neumann
type equation will be derived.

Denoting by $O^{\ddagger }$ the adjoint of an operator $O$ with respect to
the pseudo-inner product 
\begin{equation}
(\cdot ,\cdot )_{\eta }=(\cdot ,\eta \cdot )  \label{IP}
\end{equation}%
(where $(\cdot ,\cdot )$ represent the standard quaternionic inner product
in the space $\mathbb{Q}^{n}$), we have%
\begin{equation}
O^{\ddagger }=\eta ^{-1}O^{\dagger }\eta  \label{adj}
\end{equation}%
so that for any $\eta $-pseudo-Hermitian operator, i. e., satisfying the
relation,

\begin{equation}
\eta O\eta ^{-1}=O^{\dagger },  \label{pseudoh}
\end{equation}%
one has, $O=O^{\ddagger }.$

If $O$ is $\eta $-pseudo-Hermitian, Eq. (\ref{pseudoh}) immediately implies
that $\eta O$ is Hermitian, so that the expectation value of $O$ in the
state $|\psi \rangle $ with respect to the pseudo-inner product (\ref{IP})
can be obtained,%
\begin{equation}
\left\langle \psi \right| \eta O\left| \psi \right\rangle =\func{Re}\mathrm{%
Tr}(|\psi \rangle \langle \psi |\eta O)=\func{Re}\mathrm{Tr}(\tilde{\rho}O),
\label{ex}
\end{equation}%
where $\tilde{\rho}=|\psi \rangle \langle \psi |\eta $.

More generally, if $\rho $ denotes a generic quaternionic density (hence
Hermitian and positive) matrix, we can associate with it a \textit{%
generalized} density matrix $\tilde{\rho}$ by means of a one-to-one mapping
in the following way:%
\begin{equation}
\tilde{\rho}=\rho \eta  \label{rho}
\end{equation}%
and obtain $\left\langle O\right\rangle _{\eta }=\func{Re}\mathrm{Tr}(\tilde{%
\rho}O).$

Note that $\tilde{\rho}$ is $\eta $-pseudo-Hermitian: 
\begin{equation*}
\tilde{\rho}^{\dagger }=\eta \rho =\eta \tilde{\rho}\eta ^{-1}.
\end{equation*}

As in the Hermitian case \cite{compent}, \cite{Asor}, \cite{gallipoli}, \cite%
{ASS}, Eq. (\ref{ex}) immediately implies that the expectation value of an $%
\eta $-pseudo-Hermitian operator $O$ on the generalized state $\tilde{\rho}$
depends on the quaternionic parts of $O$ and $\tilde{\rho}$, only if both
the operator and the generalized state are represented by genuine
quaternionic matrices. Hence, if a $\eta $-pseudo-Hermitian operator $O$ is
described by a \emph{complex} matrix, its expectation value does not depend
on the quaternionic part $j\tilde{\rho}_{\beta }$\ of the state $\tilde{\rho}%
=\tilde{\rho}_{\alpha }+j\tilde{\rho}_{\beta }$ .

It was shown that whenever the quaternionic Hamiltonian $H$ of a quantum
system is $\eta $-pseudoanti-Hermitian, i. e.,%
\begin{equation}
\eta H\eta ^{-1}=-H^{\dagger },  \label{pseudoermiticity}
\end{equation}%
where $\eta =\eta ^{\dagger }$ , the pseudo-inner product (\ref{IP}), is
invariant under the time traslation generated by $H$ provided that $\eta $
does not depend on $t$\cite{scola}:

\begin{equation}
\left\langle \psi (t)\right| \eta \left| \psi (t)\right\rangle =\left\langle
\psi (0)\right| \eta \left| \psi (0)\right\rangle .  \label{6bis}
\end{equation}%
Denoting by $V(t)$ the evolution operator

\begin{equation}
|\psi (t)\rangle =V(t)|\psi (0)\rangle  \label{evstate}
\end{equation}%
Eq. (\ref{6bis}) immediately implies%
\begin{equation}
V^{\dagger }\eta V=\eta ,  \label{etaunitary}
\end{equation}%
i. e., $V$ is $\eta $-unitary. Whenever $H$ is time-independent, $\eta $%
-unitarity of $V$ is quite apparent, owing to its explicit form $%
V(t)=e^{-Ht} $ ($\hslash =1$) and invoking $\eta $-pseudoanti-Hermiticity of 
$H$.

Moreover, from

\begin{equation}
\rho (t)=V\rho (0)V^{\dagger }  \label{9}
\end{equation}%
by easy calculations, we obtain for a generalized quaternionic density
matrix $\tilde{\rho}$ 
\begin{equation}
\tilde{\rho}\left( t\right) =V(t)\tilde{\rho}(0)V(t)^{-1}.  \label{conserbis}
\end{equation}%
In conclusion, $\eta $-unitarity of the time-evolution is a consequence of
the $\eta $-pseudoanti-Hermiticity of $H$.

Conversely, let us assume unitarity of the time-evolution with respect to a
(possibly time-dependent) $\eta $-inner product:

\begin{equation*}
\langle \psi (0)|\eta (0)|\phi (0)\rangle =\langle \psi (t)|\eta (t)|\phi
(t)\rangle .
\end{equation*}%
Then, this condition is equivalent to

\begin{equation*}
\eta (0)=V^{\dagger }(t)\eta (t)V(t).
\end{equation*}%
Differentiating both sides of the preceding equation we immediately get

\begin{equation}
\left( \frac{d}{dt}\eta (t)\right) \eta ^{-1}(t)=H^{\dagger }(t)+\eta
(t)H(t)\eta (t)^{-1}  \label{et}
\end{equation}%
where

\begin{equation}
H(t)=-\left( \frac{d}{dt}V(t)\right) V^{-1}(t).  \label{Vevol}
\end{equation}%
Eq. (\ref{et}) shows that $H(t)$ is $\eta $-pseudoanti-Hermitian if and only
if $\eta $ is time-independent. In this case, the time evolution of $\tilde{%
\rho}(t)$ is described at finite level by Eq. (\ref{conserbis}) and at
infinitesimal level by the usual Liouville-von Neumann equation:

\begin{equation}
\frac{d}{dt}\tilde{\rho}\left( t\right) =-[H(t),\tilde{\rho}].
\label{liouville}
\end{equation}

From Eq. (\ref{conserbis}), the conservation of the $\eta $-pseudo-norm
immediately follows:

\begin{equation*}
\func{Re}\mathrm{Tr}\tilde{\rho}(t)=\func{Re}\mathrm{Tr}\tilde{\rho}(0).
\end{equation*}

From Eqs. (\ref{conserbis}), (\ref{etaunitary}) and the $\eta $%
-pseudo-Hermiticity of $\tilde{\rho}(0)$ we immediately get

\begin{equation*}
\eta \tilde{\rho}\left( t\right) \eta ^{-1}=\eta V(t)\tilde{\rho}%
(0)V^{-1}(t)\eta ^{-1}=V^{\dagger -1}(t)\eta \tilde{\rho}(0)\eta
^{-1}V^{\dagger }(t)=\tilde{\rho}^{\dagger }\left( t\right) ,
\end{equation*}%
i. e., $\tilde{\rho}\left( t\right) $ is $\eta $-pseudo-Hermitian.

\section{Quasistationary quaternionic Hamiltonian dynamics and their complex
projections}

In this section, we restrict ourselves to consider the space of quaternionic
quasi-Hermitian density matrices, that is the subclass of $\eta $%
-pseudo-Hermitian density matrices where $\eta =T^{2}$ for some nonsingular
bounded Hermitian operator $T$.

An important property of such generalized density matrices is that they are
positive definite; indeed, putting $\eta =T^{2}$ \ into Eq. (\ref{rho}),
from the positivity of $\rho $ we immediately obtain $T\tilde{\rho}%
T^{-1}=T\rho T=T\rho T^{\dagger }\geq 0$ \cite{Zhang}.

Then, the inner product (\ref{IP}) we introduced in the Hilbert space is
positive, so that all the usual requirements for a proper quantum
measurement theory can be maintained \cite{altern}, \cite{geyer}, \cite%
{indefinite}, \cite{mosta}. Hence, according to the discussion in Sec. 2,
the $\eta $-unitarity of the time-evolution implies that a time-dependent
quaternionic Hamiltonian operator $H(t)$ defines a consistent unitary
quaternionic quantum system if and only if $H(t)$ is $\eta $%
-pseudoanti-Hermitian for a time-independent positive $\eta $ operator. We
will call such a Hamiltonian $\eta $-quasistationary \cite{timdep}.

When one considers $\eta $-quasistationary quaternionic dynamics, any $\eta $%
-unitary operator $V(t)$ can be decomposed as follows:

\begin{equation}
V=T^{-1}U(t)T  \label{etaunitarydecomposition}
\end{equation}%
where $U^{\dagger }U=\mathbf{1}$.

In fact, by using Eq. (\ref{etaunitary}) and imposing unitarity, we
immediately get

\begin{equation*}
T^{-1}V^{\dagger }TTVT^{-1}=\mathbf{1}.
\end{equation*}

Recalling that $\eta $ is time-independent, we immediately obtain from Eqs. (%
\ref{etaunitarydecomposition}), (\ref{Vevol})

\begin{equation}
H(t)=T^{-1}\mathfrak{H}(t)T,  \label{Gererator}
\end{equation}%
where $\mathfrak{H}^{\dagger }(t)=-\mathfrak{H}(t)$.

Let us denote by $M(\mathbb{Q})$ and $M(\mathbb{C})$ the space of $n\times m$
quaternionic and complex matrices respectively and let $M=M_{\alpha
}+jM_{\beta }\in M(\mathbb{Q})$. We define the complex projection

\begin{equation*}
P:M(\mathbb{Q})\rightarrow M(\mathbb{C})
\end{equation*}
by the relation

\begin{equation}
P[M]=\frac{1}{2}[M-iMi]=M_{\alpha }.  \label{complexProjection}
\end{equation}

In order to discuss the complex projection of quaternionic $\eta $%
-quasistationary dynamics, we recall the following properties \cite{gius}:

\textit{i) The complex projection }$\tilde{\rho}_{\alpha }$ \textit{of a }$%
\eta $\textit{-quasi-Hermitian quaternionic matrix }$\tilde{\rho}=$ $\tilde{%
\rho}_{\alpha }+j\tilde{\rho}_{\beta }$ \textit{is }$\eta $\textit{%
-quasi-Hermitian if and only if the entries of }$\eta $ \textit{are complex;}

\textit{ii) The complex projection }$\tilde{\rho}_{\alpha }$ \textit{of a }$%
\eta $\textit{-quasi-Hermitian quaternionic matrix }$\tilde{\rho}=$ $\tilde{%
\rho}_{\alpha }+j\tilde{\rho}_{\beta }$ \textit{with a complex positive }$%
\eta $\textit{,} \textit{is positive and }$\func{Re}\mathrm{Tr}\tilde{\rho}%
_{\alpha }=1$\textit{.}

We sketch here the proof of property ii). By property i) it is $\tilde{\rho}%
_{\alpha }=\rho _{\alpha }\eta $. Since $\eta =T^{2}$, from the positivity
of $\rho _{\alpha }$ we immediately obtain

\begin{equation*}
T\tilde{\rho}_{\alpha }T^{-1}=T\rho _{\alpha }T^{\dagger }\geq 0,
\end{equation*}%
hence $\tilde{\rho}_{\alpha }\geq 0$. Furthermore, trivially, $\func{Re}%
\mathrm{Tr}\tilde{\rho}_{\alpha }=\func{Re}\mathrm{Tr}\widetilde{\rho }=1$.

Now, let us consider a dynamics ruled by a $\eta $-quasistationary
quaternionic Hamiltonian. Eq. (\ref{conserbis}) shows that this dynamics
represent a mapping into the set of $\eta $-quasi-Hermitian quaternionic
density matrices; moreover, if $\eta $ is complex and positive, properties
i) and ii) ensure that the complex projection $\tilde{\rho}_{\alpha }$ of $%
\tilde{\rho}$ is a complex $\eta $-quasi-Hermitian density matrix for any $%
\tilde{\rho}$. Hence, we can conclude that the complex projection of a $\eta 
$-quasistationary quaternionic Hamiltonian dynamics, (with $\eta $ complex
positive) is a complex stochastic dynamics in the space of $\eta $%
-quasi-Hermitian complex density matrices.

The explicit form of such dynamics can be obtained from Eqs. (\ref%
{etaunitary}-\ref{conserbis}) decomposing $V$ in terms of its complex parts $%
V_{\alpha }$ and $V_{\beta }$: $V=V_{\alpha }+jV_{\beta }$. Indeed by Eq. (%
\ref{9}) one has

\begin{equation*}
\rho _{\alpha }(0)\rightarrow \rho _{\alpha }(t)=V_{\alpha }\rho _{\alpha
}(0)V_{\alpha }^{\dagger }+V_{\beta }^{\ast }\rho _{\alpha }^{\ast
}(0)V_{\beta }^{T}+V_{\alpha }\rho _{\beta }^{\ast }(0)V_{\beta
}^{T}-V_{\beta }^{\ast }\rho _{\beta }(0)V_{\alpha }^{\dagger },
\end{equation*}%
which is a complex positive map in the space of Hermitian density matrices $%
\rho _{\alpha }$ ($\ast $ and $T$ denote as usual complex conjugation and
transposition, respectively). It follows that

\begin{equation}
\widetilde{\rho }_{\alpha }(t)=(V_{\alpha }\rho _{\alpha }(0)V_{\alpha
}^{\dagger }+V_{\beta }^{\ast }\rho _{\alpha }^{\ast }(0)V_{\beta
}^{T}+V_{\alpha }\rho _{\beta }^{\ast }(0)V_{\beta }^{T}-V_{\beta }^{\ast
}\rho _{\beta }(0)V_{\alpha }^{\dagger })\eta .  \label{finitecomppro}
\end{equation}

Eq. (\ref{finitecomppro}) can be rewritten in term of $\widetilde{\rho }%
_{\alpha }(0)$ and $\widetilde{\rho }_{\beta }(0)$ putting $V^{-1}=W_{\alpha
}+jW_{\beta }$ and using the relations $V_{\alpha }^{\dagger }\eta =\eta
W_{\alpha }$, $-V_{\beta }^{T}\eta =\eta ^{\ast }W_{\beta }$ (see Eq. (\ref%
{etaunitary}))

\begin{equation*}
\widetilde{\rho }_{\alpha }(t)=V_{\alpha }\widetilde{\rho }_{\alpha
}(0)W_{\alpha }-V_{\beta }^{\ast }\widetilde{\rho }_{\alpha }^{\ast
}(0)W_{\beta }-V_{\alpha }\widetilde{\rho }_{\beta }^{\ast }(0)W_{\beta
}-V_{\beta }^{\ast }\widetilde{\rho }_{\beta }(0)W_{\alpha }.
\end{equation*}%
At a infinitesimal level the previous equation get

\begin{equation}
\frac{d}{dt}\widetilde{\rho }_{\alpha }=-[H_{\alpha },\widetilde{\rho }%
_{\alpha }]+H_{\beta }^{\ast }\widetilde{\rho }_{\beta }-\widetilde{\rho }%
_{\beta }^{\ast }H_{\beta }  \label{gencompro}
\end{equation}%
where the symplectic decomposition of $H$ has been used.

It is worthwhile to stress that, unlike what happens whenever $H$ is
time-independent (in such case a one-parameter semigroup dynamics can always
be associated with $H$ \cite{gius}), in the general case we considered here
the evolution operator does not obey a semigroup law, as the example in the
following section will show explicitly.

\section{An example}

Let us observe firstly that the more general 2-dimensional complex positive $%
\eta $ operator is given by

\begin{equation}
\eta =T^{2}=\left( 
\begin{array}{cc}
x & z \\ 
z^{\ast } & y%
\end{array}%
\right) \left( 
\begin{array}{cc}
x & z \\ 
z^{\ast } & y%
\end{array}%
\right) =\left( 
\begin{array}{cc}
x^{2}+|z|^{2} & (x+y)z \\ 
(x+y)z^{\ast } & y^{2}+|z|^{2}%
\end{array}%
\right)  \label{eta}
\end{equation}%
where $x,y\in \mathbb{R}$, $z\in \mathbb{C}$ and $xy\neq |z|^{2}$ and

\begin{equation*}
T^{-1}=\frac{1}{xy-|z|^{2}}\left( 
\begin{array}{cc}
y & -z \\ 
-z^{\ast } & x%
\end{array}%
\right) .
\end{equation*}

The complex dynamical map we will study, is obtained as the complex
projection of a deformation of the quaternionic unitary map:

\begin{equation}
U(t)=\left( 
\begin{array}{cc}
\sqrt{1-(\sin 2t)^{4}}+je^{-i\theta }(\sin 2t)^{2} & 0 \\ 
0 & 1%
\end{array}%
\right) ,\text{ }\theta \in \mathbb{R}
\end{equation}%
to which corresponds the anti-Hermitian time-dependent Hamiltonian

\begin{equation}
\mathfrak{H}(t)=-\left( \frac{d}{dt}U(t)\right) U^{\dagger }(t)=\left( 
\begin{array}{cc}
j\frac{-4e^{-i\theta }\sin 2t\cos 2t}{|\cos 2t|\sqrt{1+(\sin 2t)^{2}}} & 0
\\ 
0 & 0%
\end{array}%
\right) .  \label{HB1}
\end{equation}%
We extensively studied such Hamiltonian \cite{ASS}, which generalizes to the
quaternionic case a complex stochastic dynamics, arising in some decoherence
modeling schemes \cite{S}.

From the previous two equations and Eqs. (\ref{Gererator}), (\ref%
{etaunitarydecomposition}) we get

\begin{equation}
H(t)=T^{-1}\mathfrak{H}(t)T=j\frac{-4e^{-i\theta }\sin 2t\cos 2t}{%
(xy-|z|^{2})|\cos 2t|\sqrt{1+(\sin 2t)^{2}}}\left( 
\begin{array}{cc}
yx & yz \\ 
-zx & -z^{2}%
\end{array}%
\right) ,  \label{Hquasi}
\end{equation}%
and

\begin{equation}
V(t)=T^{-1}U(t)T=\frac{1}{xy-|z|^{2}}\left( 
\begin{array}{cc}
yxq-|z|^{2} & y(q-1)z \\ 
-z^{\ast }(q-1)x & -z^{\ast }qz+xy%
\end{array}%
\right)  \label{Vetaunitary}
\end{equation}%
where, $q=\sqrt{1-(\sin 2t)^{4}}+je^{-i\theta }(\sin 2t)^{2}$.

Let the initial ''pure'' state be

\begin{equation}
\tilde{\rho}(0)=\frac{1}{2}\left( 
\begin{array}{cc}
1 & 0 \\ 
0 & 1%
\end{array}%
\right) +\frac{1}{2(xy-|z|^{2})}j\left( 
\begin{array}{cc}
-(x+y)z^{\ast }e^{-i\theta } & -(|z|^{2}+y^{2})e^{-i\theta } \\ 
(|z|^{2}+x^{2})e^{-i\theta } & (x+y)ze^{-i\theta }%
\end{array}%
\right) ,  \label{in}
\end{equation}%
according with Eq. (\ref{conserbis}) the final state reads

\begin{eqnarray}
\tilde{\rho}(t) &=&\frac{1}{2}\left( 
\begin{array}{cc}
1 & 0 \\ 
0 & 1%
\end{array}%
\right) +\frac{(\sin 2t)^{2}}{2(xy-|z|^{2})}\left( 
\begin{array}{cc}
yz^{\ast }-xz & y^{2}-z^{2} \\ 
(x^{2}-z^{\ast 2}) & -yz^{\ast }+xz%
\end{array}%
\right) +  \notag \\
&&j\frac{e^{-i\theta }\sqrt{1-(\sin 2t)^{4}}}{2(xy-|z|^{2})}\left( 
\begin{array}{cc}
-z^{\ast }(x+y) & -|z|^{2}-y^{2} \\ 
|z|^{2}+x^{2} & z(x+y)%
\end{array}%
\right) .
\end{eqnarray}

The complex projection stochastic dynamics is given by

\begin{equation*}
\frac{1}{2}\left( 
\begin{array}{cc}
1 & 0 \\ 
0 & 1%
\end{array}%
\right) \rightarrow \frac{1}{2}\left( 
\begin{array}{cc}
1 & 0 \\ 
0 & 1%
\end{array}%
\right) +\frac{(\sin 2t)^{2}}{2(xy-|z|^{2})}\left( 
\begin{array}{cc}
yz^{\ast }-xz & y^{2}-z^{2} \\ 
(x^{2}-z^{\ast 2}) & -yz^{\ast }+xz%
\end{array}%
\right) .
\end{equation*}

Note that the semigroup composition law does not hold. In fact, by a direct
computation it is easy to verify that

\begin{equation*}
P[V(t)\tilde{\rho}(0)V^{-1}(t)]\neq P[V(t-t^{\prime })V(t^{\prime })\tilde{%
\rho}(0)V^{-1}(t^{\prime })V^{-1}(t-t^{\prime })],
\end{equation*}%
indeed,%
\begin{equation*}
P[(V(t)\tilde{\rho}(0)V^{-1}(t))_{21}]=(\sin 2t)^{2}
\end{equation*}%
while

\begin{eqnarray*}
&&P[(V(t-t^{\prime })V(t^{\prime })\tilde{\rho}(0)V^{-1}(t^{\prime
})V^{-1}(t-t^{\prime }))_{21}] \\
&=&(\cos 2(t-t^{\prime }))^{2}(\sin 2t^{\prime })^{2}-[(1-(\cos
2(t-t^{\prime }))^{4})(1-(\sin 2t^{\prime })^{4})]^{\frac{1}{2}}.
\end{eqnarray*}

\section{Final remark}

Let a Hermitian nonsingular quaternionic operator $\eta $ be given. Then,
the more general $\eta $-pseudoanti-Hermitian quaternionic Hamiltonian $H$
can be written in the following way \cite{gius}:

\begin{equation}
H=F\eta ,  \label{fact}
\end{equation}%
where $F^{\dagger }=-F$.

This peculiarity can be useful to obtain the full class of $\eta $%
-quasistationary quaternionic dynamics. In fact, for any time-dependent
anti-Hermitian quaternionic operator $F(t)$, we can construct a
corresponding anti-Hermitian operator $\mathfrak{H}(t)=\eta ^{\frac{1}{2}%
}F(t)\eta ^{\frac{1}{2}}$, and from Eq. (\ref{Gererator}) we can state that

\begin{equation}
H(t)=\eta ^{-\frac{1}{2}}\mathfrak{H}(t)\eta ^{\frac{1}{2}},  \label{f}
\end{equation}%
is a $\eta $-quasistationary Hamiltonian, and construct the stochastic
(complex) map associated with it, by the methods used in the example above.
Conversely, let a $\eta $-quasistationary quaternionic dynamics be given,
then, a time-dependent anti-Hermitian operator $F(t)$ can be associated with
it.

\textbf{Acknowledgements }One of us, G. S., wishes to thank prof. M. Znojil
for useful discussions and financial support during a short stay in Prague.


\bigskip

\bigskip

\begin{thebibliography}{99}
\bibitem{proc} Proceedings of the Ist, IInd, IIIrd, IVth and Vth
International Workshops on \textit{''Pseudo-Hermitian Hamiltonians in
Quantum Physics''} in \textit{Czech. J. Phys.} \textbf{54} (2004), Nos. 1,
10, \textit{Czech. J. Phys.} \textbf{55} (2005), \textit{J. Phys. A }\textbf{%
39 }(2006) and \textit{Czech. J. Phys.} \textbf{56} (2006), respectively.

\bibitem{mosta} A. Mostafazadeh, \textit{J. Phys. A} \textbf{36} (2003) 7081.

\bibitem{bologna} G. Scolarici and L. Solombrino, \textit{Czech. J. Phys. }%
\textbf{56} (2006) 935.

\bibitem{compent} G. Scolarici and L. Solombrino ``Complex entanglement and
quaternionic separability'' in \textit{The Foundations of Quantum Mechanics:
Historical Analysis and Open Questions-Cesena 2004}, C. Garola, A. Rossi and
S. Sozzo eds. (World Scientific, Singapore, 2006).

\bibitem{Asor} M. Asorey and G. Scolarici, \textit{J. Phys. A }\textbf{39}
(2006) 9727.

\bibitem{gallipoli} M. Asorey, G. Scolarici and L. Solombrino, \textit{%
Theor. Math. Phys. }\textbf{151} (2007) 733.

\bibitem{ASS} M. Asorey, G. Scolarici and L. Solombrino, \textit{Phys. Rev.
A }\textbf{76} (2007) 12111.

\bibitem{scola} G. Scolarici, \textit{J. Phys. A }\textbf{35} (2002) 7493.

\bibitem{gius} G. Scolarici, \textit{SIGMA} \textbf{3} (2007) 088.

\bibitem{timdep} A. Mostafazadeh, \textit{Phys. Lett. B} \textbf{650} (2007)
208.

\bibitem{s} P. M. Mathews, J. Rau and E. C. G. Sudarshan, \textit{Phys. Rev. 
}\textbf{121} (1961) 920.

\bibitem{Zhang} F. Zhang, \textit{Lin. Alg. Appl. }\textbf{251} (1997) 21.

\bibitem{altern} A. Blasi, G. Scolarici and L. Solombrino, \textit{J. Math.
Phys. }\textbf{46} (2005) 42104.

\bibitem{geyer} F. G. Sholtz, H. B. Geyer and F. J. W. Hahne, \textit{Ann.
Phys. }\textbf{213} (1992) 74.

\bibitem{indefinite} A. Blasi, G. Scolarici and L. Solombrino, \textit{J.
Phys. A} \textbf{37} (2004) 4335.

\bibitem{S} C. A. Rodriguez, A. Shaji and E. C. G. Sudarshan ''Dynamics of
Two Qubits: Decoherence and Entanglement Optimization Protocol'' arXiv:
quant-ph/0504051.
\end{thebibliography}
\end{document}